\documentclass{aastex63}

\usepackage{todonotes}
\usepackage{amsmath}
\usepackage{bm}
\usepackage{xcolor}
\usepackage{CJK}

\received{2022 September 28}
\revised{2023 February 3}
\accepted{2023 February 6}
\published{2023 March 7}

\shorttitle{Desi Provabgs Emulator}
\shortauthors{Kwon, Hahn \& Alsing}

\graphicspath{{./}{figures/}}

\begin{document}

\title{Neural Stellar Population Synthesis Emulator for the DESI PROVABGS}
\email{jameskwon@ucsb.edu}

\author{K.J. Kwon}
\affiliation{Department of Physics, University of California, Santa Barbara, Broida Hall, Santa Barbara, CA 93106, USA}
\author{ChangHoon Hahn}
\affiliation{Department of Astrophysical Sciences, Princeton University, Peyton Hall, Princeton, NJ 08544, USA}
\author{Justin Alsing}
\affiliation{Oskar Klein Centre for Cosmoparticle Physics, Department of Physics, Stockholm University, Stockholm SE-106 91, Sweden}

\begin{abstract}
The Probabilistic Value-Added Bright Galaxy Survey (PROVABGS) catalog will provide the posterior distributions of physical properties of $>10$ million DESI Bright Galaxy Survey (BGS) galaxies.
Each posterior distribution will be inferred from joint Bayesian modeling of observed photometry and spectroscopy using Markov Chain Monte Carlo sampling and the \cite{provabgs} stellar population synthesis (SPS) model. 
To make this computationally feasible, PROVABGS will use a neural emulator for the SPS model to accelerate the posterior inference. 
In this work, we present how we construct the emulator using the \cite{speculator}~approach and verify that it can be used to accurately infer galaxy properties. 
We confirm that the emulator is in excellent agreement with the original SPS model with $\ll 1\%$ error and is $100\times$ faster. 
In addition,  
we demonstrate that the posteriors of galaxy properties derived using the emulator are also in excellent agreement with those inferred using the original model. 
The neural emulator presented in this work is essential in bypassing the computational challenge posed in constructing the PROVABGS catalog. Furthermore, it demonstrates the advantages of emulation for scaling sophisticated analyses to millions of galaxies. 
\end{abstract}

\section{Introduction} \label{sec:intro}
Over the next decade, spectroscopic galaxy surveys such as the 
Dark Energy Spectroscopic Instrument~\citep[DESI;][]{desicollaboration2016, desi_instrument}, 
the Prime  Focus Spectrograph~\citep[PFS;][]{takada2014}, 
William Herschel Telescope Enhanced Area Velocity Explorer~\citep[WEAVE;][]{weave}, 
and the 4-metre Multi-Object Spectroscopic Telescope~\citep[4MOST;][]{4most}
will measure the spectra of millions of galaxies. 
They will provide statistically powerful samples of galaxies at 
higher redshifts, low stellar mass ranges, and other less explored 
regimes.
The DESI Bright Galaxy Survey~\citep[BGS;][]{hahn2022}, 
for instance,  will provide a magnitude-limited galaxy sample out to $z\sim0.6$ and 
probe dwarf galaxies down to $\sim10^7M_\odot$ in the local universe.
From the spectra of galaxies in these new regimes, we can measure their
detailed physical properties and shed light on their formation and 
evolution. 

We currently measure the physical properties of a galaxy, such as its stellar mass ($M_*$), star formation rate (SFR), metallicity ($Z$), 
and age ($t_{\rm age}$), by modeling its spectral energy distribution  
(SED), or spectra, using stellar population synthesis~\citep[SPS; \emph{e.g.}][see \citealt{walcher2011, conroy2013} 
for a comprehensive review]{bruzual2003, maraston2005, conroy2009}.
Theoretical SPS models are compared to observed spectra using Bayesian
inference to accurately quantify parameter uncertainties and 
degeneracies~\citep{acquaviva2011,
chevallard2016, leja2017, carnall2018, johnson2021}. 
The Bayesian approach enables marginalization over nuisance 
parameters that model the effects of observational systematics 
(\emph{e.g.} flux calibration).
Posteriors of galaxies in a population can also be combined using a Bayesian 
hierarchical approach for more accurate population  inference~\citep[\emph{e.g.}][]{leja2019a, leja2021b, alsing2022, leistedt2022}.
However, one major limitation of current Bayesian SED modeling methods 
is the computational resources that they require.  
Current methods, which use Markov Chain Monte Carlo (MCMC) techniques to 
sample the posterior, take 10--100 CPU hours per galaxy~\citep[\emph{e.g.}][]{carnall2019a, tacchella2021}. 
This makes it prohibitive to apply state-of-the-art galaxy SED modeling to
the next-generation galaxy surveys. 

\cite{speculator} recently demonstrated that neural emulators can be used
to accelerate SED model evaluations by three to four orders of magnitude.
With the speed-up of neural emulators, posterior inference can be reduced 
from hours to minutes per galaxy.
In this work, we present a neural emulator specifically designed for the 
SPS model that will be used to construct the PRObabilistic Value-Added 
Bright Galaxy Survey~\citep[PROVABGS\footnote{\href{https://changhoonhahn.github.io/provabgs/current/}{https://changhoonhahn.github.io/provabgs}}]{provabgs} catalog. 
PROVABGS will provide the posterior distributions of physical properties 
of $>10$ million DESI BGS galaxies from joint SED modeling of their optical 
spectrophotometry.
Its SED model is based on an SPS model with a non-parametric star formation 
history (SFH), a non-parametric metallicity history (ZH) that varies with time,
and a flexible dust attenuation prescription.
The goal of this work is to demonstrate that our emulator accurately 
reproduces the SED model (with  $\ll1\%$ errors) as well as the posterior 
on galaxy properties.

We begin in Section~\ref{sec:sps} with a brief description of the PROVABGS
SED model. 
We then present how we design and train our neural emulator in Section~\ref{sec:emulator}. 
In Section~\ref{sec:result}, we validate the emulator.

\section{Stellar Population Synthesis}\label{sec:sps}
The PROVABGS catalog will use the state-of-the-art \cite{provabgs} SPS model to infer the posteriors of the physical properties of BGS galaxies. In this section, we briefly introduce the PROVABGS~SPS model that we emulate. In the PROVABGS SPS model, a galaxy SED is modeled as a combination of the single stellar populations (SSPs), a group of stars of the same age and metallicity, formed over the lifetime of the galaxy. 

The PROVABGS~SPS model uses non-parametric prescriptions for its
SFH and ZH. 
They are treated as linear combinations of basis functions derived from applying 
non-negative matrix factorization~(NMF;~\citealt{nmf1, nmf2, nmf3}) to the SFHs 
and ZHs of simulated galaxies in the \textsc{Illustris} cosmological 
hydrodynamic simulation~\citep{illustris2, illustris1}. 
For SFH, the model uses four NMF bases with coefficients, $\beta_1, \beta_2, \beta_3, 
\beta_4$, and include an SSP starburst component parameterized with 
$f_{\text{burst}}$ and $t_{\text{burst}}$ for additional stochasticity. 
$f_{\text{burst}}$ denotes the fraction of the total stellar mass formed during the starburst and $t_{\text{burst}}$ is the time of the starburst event.  
For the ZH that varies with time, the model uses two NMF bases with coefficients $\gamma_1$ and $\gamma_2$.

From the SFH and ZH, we calculate the rest-frame luminosity of a galaxy. 
The SFH and ZH are binned into logarithmically-spaced lookback time 
($t_{\text{lookback}}$) that are truncated at the age of the galaxy. 
Then, for each $t_{\text{lookback}}$ bin, the luminosity of an SSP is computed,
based on the metallicity in the bin, using   
\texttt{Flexible Stellar Population Synthesis}\footnote{we use \texttt{Python} wrapper of \texttt{FSPS}, \texttt{Python-Fsps} \citep{python_fsps}}\textsuperscript{,}\footnote{\href{https://github.com/dfm/python-fsps}{https://github.com/dfm/python-fsps}}(hereafter \texttt{FSPS};~\citealt{conroy2009, fsps_model})
with MIST isochrones (\citealt{mist_3.1, mist_3.2, mist_3.3, mist_2, mist_1}), the~\cite{chabrier_2003} initial mass function, MILES stellar library~\citep{miles} for the wavelengths between 3800--7100\text{\AA} and BaSeL library \citep{basel1, basel2, basel3} outside of that range. 
The SSP luminosity is attenuated using the two-component dust attenuation by~\cite{dust_charlot}. 
SSPs younger than 100 Myr are first attenuated by birth clouds. 
Afterward, all SSPs are attenuated by the~\cite{kriek_conroy} attenuation curve.
The SSP luminosities are then normalized by the total stellar mass formed
in their $t_{\rm lookback}$ bins and then combined to obtain the dust-attenuated SED. 

In summary, apart from nuisance parameters, the PROVABGS~SPS model 
has 12 model parameters: total stellar mass, four NMF
SFH basis coefficients, two starburst components, two NMF ZH basis 
coefficients, and three dust attenuation parameters. 
For the full list of the PROVABGS model parameters and more details on 
the PROVABGS~SPS model, we refer readers to \cite{provabgs}. 

\section{PROVABGS Emulator}\label{sec:emulator}
The PROVABGS catalog will infer $>10$ million posteriors using MCMC sampling.
As we later demonstrate, accurately estimating the posterior requires $\gtrsim 120$,000 SPS model evaluations, given the dimensionality of our SPS model.
If we evaluate the PROVABGS SPS model using \texttt{FSPS}, it requires $>10$ CPU hours per galaxy and makes PROVABGS computationally infeasible (we will refer to the PROVABGS SPS model evaluated using \texttt{FSPS} as PROVABGS-FSPS). 
We, therefore, construct a neural emulator for the SPS model that will accelerate the model evaluations. 
In this section, we describe how we construct the emulator.  

We follow the neural emulation framework from~\cite{speculator}. 
The emulator consists of a neural network that takes SPS model parameters as input and predicts 
coefficients of a fixed set of $N_{\rm PCA}$ Principal Component Analysis (PCA) basis vectors for 
galaxy SEDs. 
The PCA coefficients and basis vectors are then linearly combined to predict the model SED as a function of wavelength. 
In \cite{speculator}, they use a single emulator over the full SED wavelength coverage. 
In this work, however, we divide the wavelength coverage into four contiguous ranges, 2000--3600, 
3600--5500, 5500--7410, and 7410--60000\AA, to achieve higher accuracy. 
The wavelength ranges have 127, 2109, 2113, and 549 spectral bins, respectively.
We also use separate emulators for the NMF and burst contributions to the SFH. 
In total, we use 8 emulators. 
For each emulator, we use $N_{\text{PCA}}=\{50$, $50$, $50$, $30\}$ PCA components, which 
we determined through experimentation to achieve $\ll1$\% PCA reconstruction error throughout 
the entire wavelength. 
For the neural networks, we use the same fully-connected architecture and activation 
function as \cite{speculator}. 
We determine the number of hidden layers and units through experimentation:
8 hidden layers and 256 units for the NMF emulators and 6 hidden layers and 512 units for the 
burst emulators.
Our architecture is deeper than the emulator in~\cite{speculator}, because we have a more complex SPS model. 

To train the emulators, we sample 1 million parameters from the priors in  
Table~1 of \cite{provabgs} and use PROVABGS-FSPS to generate a model SED for each parameter.
We then split the dataset into a training and validation set using a 90/10 split. 
From the training set, we derive the $N_{\rm PCA}$ basis vectors and use them to transform 
the model SEDs to PCA coefficients. 
Before training, the weights of the neural network are randomly initialized from a normal distribution 
and biases are initially set to zero.
Afterward, we train the neural network using the mean squared error between the 
predicted and true PCA coefficients as the loss function with the \textsc{Adam} 
optimizer~\citep{adam}. 
The initial learning rate is set to $10^{-3}$. 
If the loss evaluated on the validation set does not improve after 20 consecutive epochs, the 
learning rate is gradually decreased following the sequence: $10^{-3},\ 5\times10^{-4},\ 10^{-4},\ldots, \ 5\times 10^{-6},\  10^{-6}$. 
After the learning rate reaches $10^{-6}$, we terminate the training when the validation 
loss does not improve for 20 consecutive epochs. 

\begin{figure*}
\centering
\includegraphics[width=\textwidth]{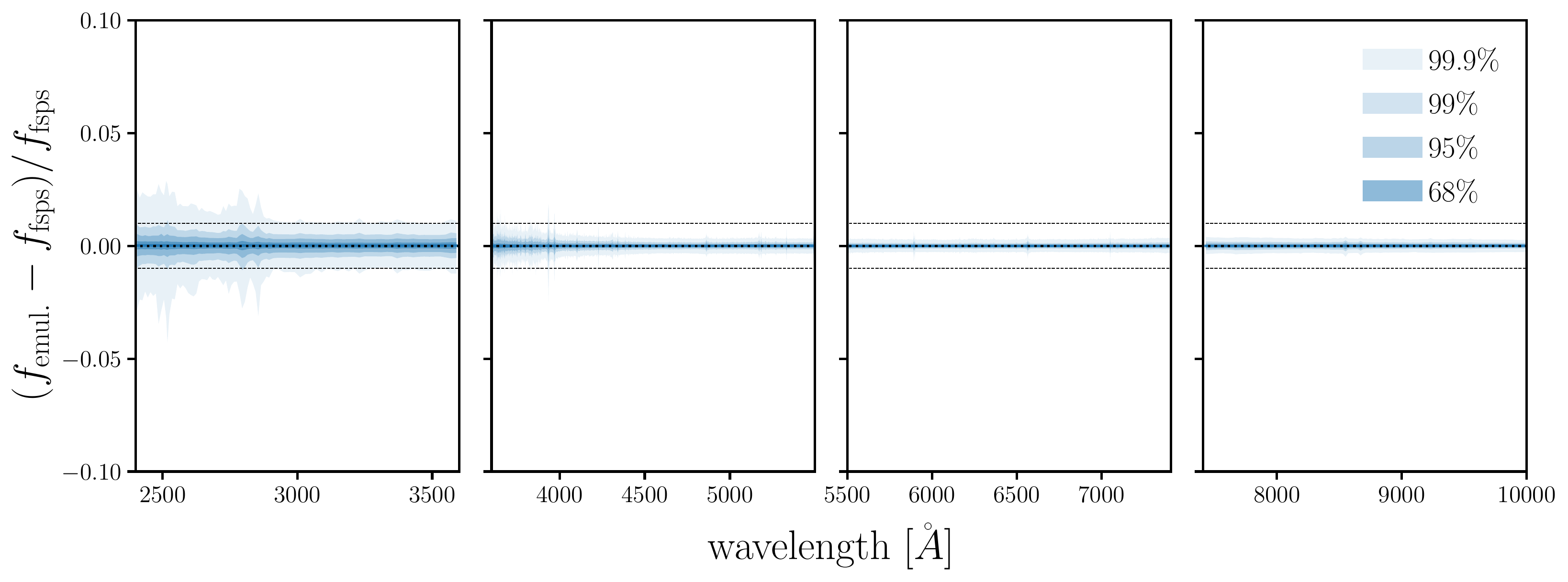}
\caption{Fractional error of the PROVABGS emulator based on the 100,000 test SEDs. The fractional errors are calculated with respect to the model SEDs calculated using PROVABGS-FSPS. Each panel represents a separate wavelength bin with a unique emulator. The fractional errors for 68, 95, 99, and 99.9\% of the test SEDs are shown with different shades. The dashed lines represent $\pm1\%$ errors. 
The emulator can reconstruct SEDs to sub-percent accuracy for $95$\% of the test SEDs over the full wavelength range, and for 99.9\% of the test SEDs over $\lambda > 3000\text{\AA}$. Although there is a relatively large error on $\lambda < 3000\text{\AA}$, we find that this does not impact the accuracy of the posteriors of the properties of forward-modeled BGS-like spectra, because BGS spectra generally have low SNR at these wavelengths.}
\label{fig:validation_err}
\end{figure*}
\newpage
\section{Results}\label{sec:result}
\subsection{Validating Emulator SED}\label{subsec:vali_sed}
In this section, we demonstrate that the emulator accurately reproduces the SEDs computed by PROVABGS-FSPS. 
To test the emulator, we sample 100,000 parameters from the priors and calculate their SEDs using the emulator and PROVABGS-FSPS. 
Then, we compute the fractional difference of the SEDs: $({f_{\text{emul.}}-f_{\text{fsps}}})/f_{\text{fsps}}$, where $f_{\rm emul.}$ and $f_{\rm fsps}$ denotes the emulator and PROVABGS-FSPS SED fluxes, respectively.
In Figure~\ref{fig:validation_err}, we present the fractional errors of the emulator in each of the four wavelength bins.
We mark the percentiles of the emulator reconstruction error with different shades (68, 95, 99, 99.9\%, from the darkest to lightest). 
For $\sim95$\% of the test SEDs, the fractional error is consistently within $< 1\%$ in all wavelength bins.
In fact, the SEDs from the emulator and PROVABGS-FSPS agree to a sub-percent level for $99.9$\% of the test SEDs 
in all but the first wavelength bin.
One may be concerned regarding relatively higher fractional errors at $\lambda < 3000 \text{\AA}$; however, we demonstrate in Section \ref{subsec:vali} that this does not significantly impact the accuracy of the posteriors because BGS spectra have lower signal-to-noise ratios (SNRs) near-UV wavelengths~\citep{hahn2022}.
Also, SPS modeling has larger uncertainties in the near-UV due to factors such as metallicities, stellar age, and dust \citep{conroy2013, Leja+_19, johnson2021}, which alleviates potential concerns. 
We also note that since we train separate emulators for the different wavelength bins, 
the SEDs predicted by our emulator can be discontinuous. 
However, given the accuracy of our emulator, this discontinuity does not have a significant impact. 
\begin{figure*}
\centering
\includegraphics[width=0.45\textwidth]{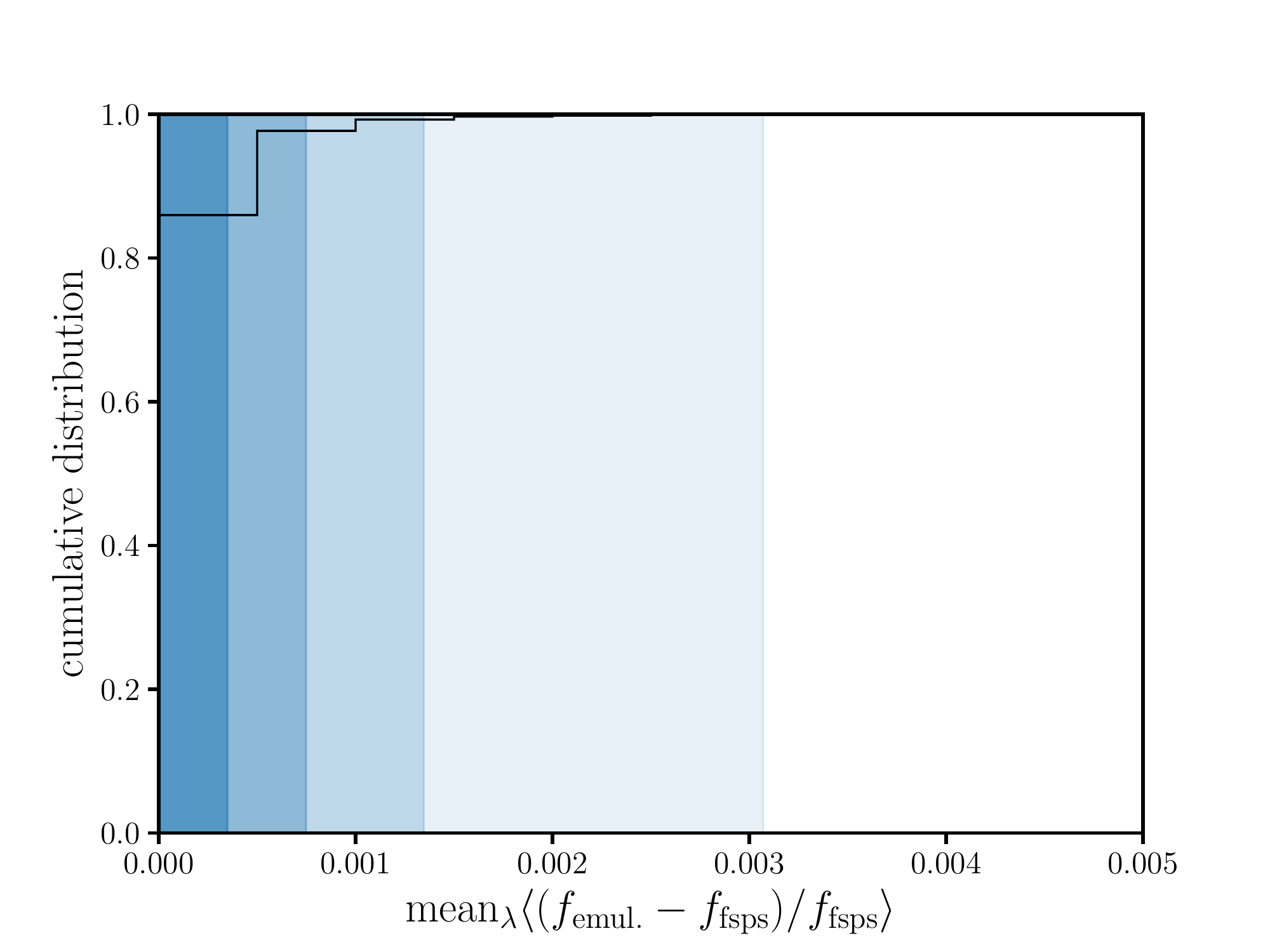}
\caption{Cumulative distribution of the absolute fractional error of the PROVABGS emulator, averaged over the wavelength $2400 < \lambda < 10000$\AA.  
Each shaded region marks the percentiles of the emulator reconstruction errors (68, 95, 99, 99.9\%, from darkest to lightest). 
The emulator produces the mean absolute fractional error $\ll$1\% on 99.9\% test SEDs.}
\label{fig:cumulative_err}
\end{figure*}
As an additional illustration of the emulator's accuracy, we present the cumulative 
distribution of the absolute fractional error averaged over the wavelength 
$2400 < \lambda < 10000$\AA~in Figure~\ref{fig:cumulative_err}. 
The shaded regions correspond to the same percentiles as Figure~\ref{fig:validation_err}. 
The emulator produces the mean fractional error $\ll 1\%$ on $>99$.9\% of the test SEDs.

\subsection{Validating Emulator Posterior\label{subsec:vali}}
\begin{figure*}
\centering
\includegraphics[width=0.6\textwidth]{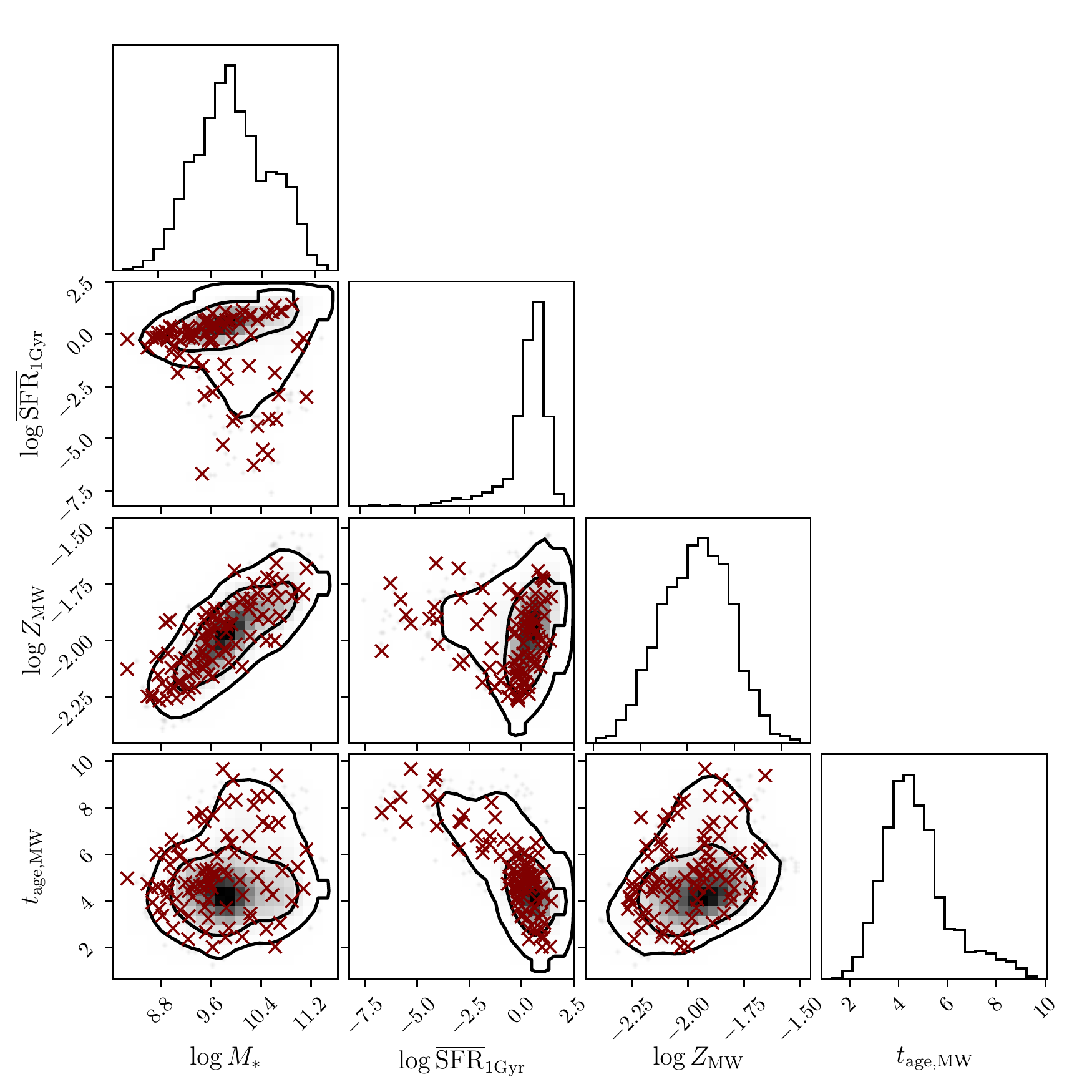}
\caption{The distribution of the physical properties of 93 \textsc{Lgal} simulated galaxies
that we use for our emulator versus PROVABGS-FSPS posterior comparison (red). 
We focus on the total stellar mass, average star formation rate over 1 Gyr, 
mass-weighted metallicity, and mass-weighted stellar age.
The black contours mark 68 and 95 percentiles of each property for the entire \textsc{Lgal} samples. 
The selected {\sc Lgal} galaxies are a representative sample of the full {\sc Lgal} sample.
We infer the posteriors of galaxy properties using both our emulator and the PROVABGS-FSPS 
from realistic BGS-like spectra constructed from these simulated galaxies. 
}\label{fig:param_distribution}
\end{figure*}

Next, we validate the posteriors of galaxy properties derived using the PROVABGS emulator 
against posteriors derived using PROVABGS-FSPS. 
We infer the posteriors from synthetic observations of BGS-like spectra from 
\cite{provabgs}. 
The synthetic spectra are constructed using $\sim2$,000 mock galaxies from the 
\textsc{L-galaxies} semi-analytic galaxy formation model (\textsc{Lgal}, hereafter;~\citealt{lgal}). 
They are calculated with \texttt{FSPS} using the SFH and ZH of the simulated galaxies. 
Then, velocity dispersion, dust attenuation, realistic noise model, and DESI BGS instrumental
effect are applied to the spectra to produce realistic DESI-like spectra. 
We refer readers to~\cite{provabgs} for the details about the forward modeling. 
Out of the $\sim2$,000 {\sc Lgal} galaxies, we select a representative subset of 93 galaxies.
In Figure~\ref{fig:param_distribution}, we present the distribution of the physical properties of these galaxies (red).
We focus on the following galaxy properties: total stellar mass ($M_*$), 
star formation rate averaged over past 1 Gyr ($\overline{\text{SFR}}_{\text{1Gyr}}$), 
mass-weighted metallicity ($Z_{\text{MW}}$), and mass-weighted age ($t_{\text{age,MW}}$).
The galaxy properties are computed in the same way as in \cite{provabgs}. 
For reference, we present the galaxy property distribution of the full {\sc Lgal} samples in black contours 
and mark the 68 and 95${\rm th}$ percentiles. 
The selected 93 samples span the full distribution of \textsc{Lgal} galaxy properties. 

We infer the posteriors of the 93 simulated galaxies from their spectra using a 
Gaussian likelihood, priors from \cite{provabgs}, and  MCMC sampling. 
For our MCMC sampler, we use the \textsc{Zeus} ensemble slice sampler \citep{zeus2,zeus1}.
Using PROVABGS-FSPS to compute SEDs, we initialize 30 walkers. 
We discard the first 1000 iterations as burn-in for 89 of the galaxies. 
For the other four galaxies, we use 6000 iterations as burn-in due to inaccurate initialization. 
The 3000 iterations after burn-in are used as samples of the posterior distributions. 
To infer the posteriors using the emulator, we initialize the walkers at the same positions as the 
walkers in the PROVABGS-FSPS chains after burn-in. 
This is to remove any impact of initialization when we later compare the two sets of posterior distributions.
We confirm that our posterior samples converge within 3,000 iterations after burn-in using 
visual inspection of the distributions of walkers and by examining the autocorrelation time.  
For the latter, we use integrated autocorrelation time (IAT), which is the number of iterations needed to sample
two points that are independent of each other. 
If $N_{\text{iteration}} > K\tau/N_{\text{walkers}}$, where $\tau$ is the converged IAT estimate
and $K$ is a constant, the walkers sampled a sufficient number of independent points to capture
posterior distributions. 
We take $K\sim50$ and calculate the mean of $\tau$ for each walker at every 50 iterations following 
\cite{gw}. 
With $N_{\text{walkers}}=30$, we obtain $\tau\approx 1000$. 
Hence, $N_{\text{iteration}}\approx 3000$ is sufficient.

\begin{figure*}
\centering
 \includegraphics[width=0.6\textwidth]{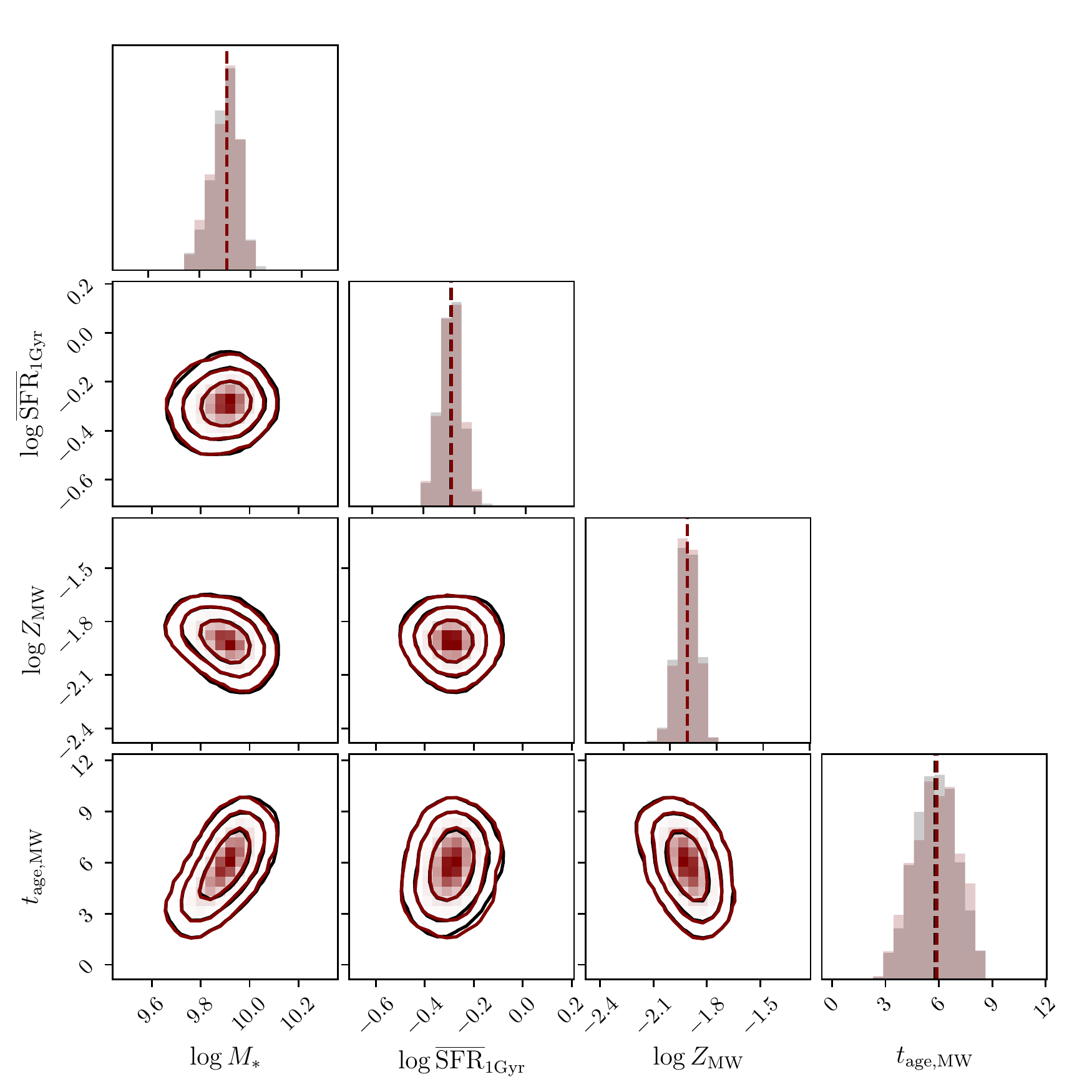}
\caption{Posterior distributions of galaxy properties derived from a synthetic BGS-like spectrum inferred 
using the emulator (red) and PROVABGS-FSPS (black). For a typical BGS-like spectrum, see \cite{provabgs}. 
The contours mark 68, 95, and 99.7 percentiles and the vertical dashed lines in the histograms denote the median of 1D marginalized posteriors. 
The posteriors inferred using the emulator and PROVABGS-FSPS are in excellent agreement.}
\label{fig:igal4_corner}
\end{figure*}

\begin{figure*}
\centering
\includegraphics[width=\textwidth]{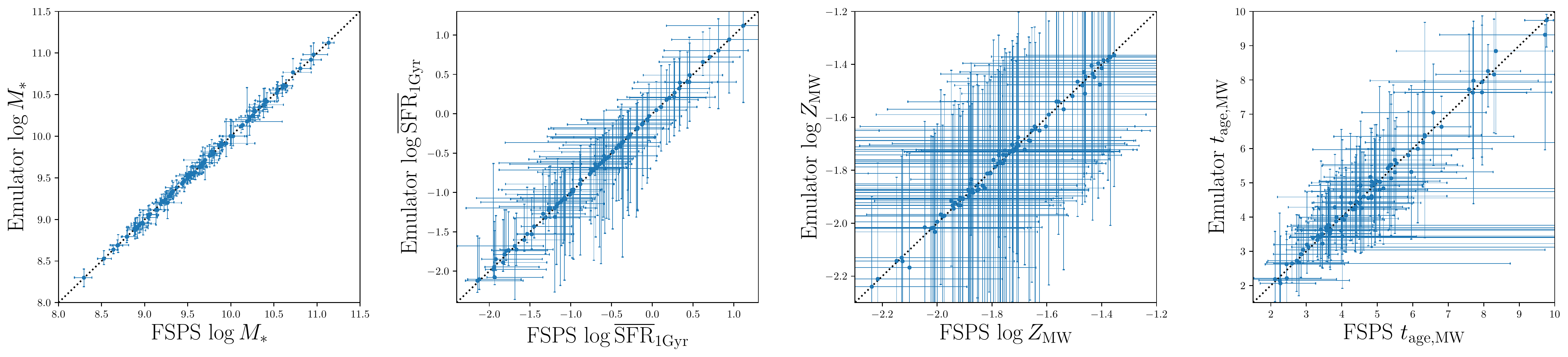}
\caption{Comparison of galaxy properties inferred using the emulator and PROVABGS-FSPS. From the left to right panels, we compare the total stellar mass ($\log M_*$), star formation rate averaged over past 1 Gyr ($\log \overline{\text{SFR}}_{\text{1Gyr}}$), mass-weighted metallicity ($\log Z_{\text{MW}}$), mass-weighted age ($t_{\text{age,MW}}$). These galaxy properties are computed using the PROVABGS SPS model based on the inferred posteriors of PROVABGS model parameters. We present the medians of inferred posteriors in the scatter points and mark the 1-$\sigma$ range with the error bars. For all samples, the posterior medians derived using the emulator agree with PROVABGS-FSPS within $\sigma_{\text{fsps}}$, and $\sigma_{\text{emul.}}$ is consistent with $\sigma_{\text{fsps}}$. As illustrated, there is an excellent agreement between the posteriors derived using the emulator and PROVABGS-FSPS.}
\label{fig:posterior}
\end{figure*}

Next, we compare the posteriors inferred using the emulator and PROVABGS-FSPS. 
From the posteriors of SPS parameters, we derive posteriors of galaxy properties $M_*$, $\overline{\text{SFR}}_{\text{1Gyr}}$, $Z_{\text{MW}}$, and $t_{\text{age,MW}}$. 
In Figure~\ref{fig:igal4_corner}, we present the emulator (red) and PROVABGS-FSPS (black) posteriors inferred from the same simulated galaxy. 
The contours represent the 68, 95, and 99.7 percentiles. The medians of the marginalized 1D posteriors are marked with vertical dashed lines in the histograms. 
The posteriors derived using the emulator and PROVABGS-FSPS are in excellent agreement. 
We also compare the medians and $1\sigma$ uncertainties of the 1D marginalized posteriors from the emulator
and PROVABGS-FSPS for the selected {\sc LGal} galaxies in Figure~\ref{fig:posterior}. 
From the leftmost panel, we compare $\log M_*$, $\log \overline{\text{SFR}}_{\text{1Gyr}}$, $\log Z_{\text{MW}}$, $\log t_{\text{age,MW}}$. 
There is excellent agreement on all 93 \textsc{Lgal} galaxies in the posterior medians and uncertainties: 
we find $<0.02\sigma$ discrepancies between the medians of the emulator and PROVABGS-FSPS posteriors. 


As the main motivation for the emulator to evaluate the PROVABGS SPS model faster, we profile the computational advantage of the emulator. 
We randomly sample 1,000 test parameters from priors and profile the elapsed time for the emulator and PROVABGS-FSPS to calculate all 1,000 SEDs. 
On a 2.6 GHz 6-Core Intel Core i7 CPU, the emulator took an average of 2.3 ms per evaluation, 
while PROVABGS-FSPS took 240 ms per evaluation.
Thus, to sample a posterior for a single galaxy, which requires $\sim 120$,000 model evaluations, the 
emulator takes $\sim 4$ minutes while PROVABGS-FSPS takes $\sim 10$ hours. 
The emulator is $>100\times$ faster than the original SPS model. 

\section{Conclusion}
We presented the neural emulator for \cite{provabgs} SPS model that will be used to 
construct the DESI PROVABGS catalog.
The emulator is constructed using the neural emulation framework of \cite{speculator}.
It consists of a fully-connected neural network that takes SPS model parameters as input
and predicts PCA coefficients, which are combined with PCA basis vectors to predict the
model SED. 
The neural network and PCA basis vectors are trained using a training set of 1 million
SPS parameters and model SEDs.
We use a similar neural network as the emulators in \cite{speculator} and determine their
architecture (number of hidden layers and units) through experimentation.

We validate the emulator using a test dataset of 100,000 model SEDs and confirm that the 
emulator reconstructs the SEDs computed using PROVABGS-FSPS with $\ll 1\%$ error for $\lambda > 3000$\AA.
The emulator reproduces the original SPS model with mean absolute fraction errors of  
$< 0.005$ on 99.9\% of the test SEDs. 
As further validation of our emulator, we compare posteriors of galaxy properties derived 
using the emulator and PROVABGS-FSPS.
The posteriors are inferred from Bayesian SED modeling of synthetic BGS-like spectra 
constructed from 93 simulated galaxies in the \textsc{L-galaxies} semi-analytic model. 
Based on autocorrelation time analysis, the MCMC sampling converges within 3,000 iterations
after burn-in. 
We find excellent agreement between the emulator and PROVABGS-FSPS posteriors. 
The medians of the posteriors deviate by $<0.02\sigma$ and are in good agreement within
the uncertainties. 
We also compare the computational cost of the emulator and PROVABGS-FSPS. 
The emulator takes 2--3 ms to model a single SED while the PROVABGS-FSPS requires 300--400 ms.
To derive a posterior for a single galaxy, our emulator requires $100\times$ less CPU time,
from 10--13 hours to 4--6 minutes. 
Additional speed-up will be achievable using GPUs \citep[\emph{e.g.}][]{alsing2022} and with
more parallel MCMC methods (\emph{e.g.} 
$\mathtt{affine}$\footnote{\href{https://github.com/justinalsing/affine}{https://github.com/justinalsing/affine}}). For reproducibility, the codebase for the training, validation, and benchmarking of the emulator is publicly available.\footnote{\href{https://github.com/changhoonhahn/gqp_mc}{https://github.com/changhoonhahn/gqp\_mc}\\
\href{https://github.com/changhoonhahn/provabgs}{https://github.com/changhoonhahn/provabgs}}

Our PROVABGS emulator is an accurate and $100\times$ faster alternative to direct SPS 
computation using PROVABGS-FSPS.
It will be essential for constructing PROVABGS and providing posteriors of galaxy properties for 
$>10$ million BGS galaxies. 
More broadly, neural emulation offers a way to scale rigorous Bayesian inference 
to the tens of millions of galaxies that will be observed by upcoming galaxy surveys
(\emph{e.g.} DESI, PFS, WEAVE, 4MOST, and Rubin).
Applications of neural emulation, such as the one presented in this work, will enable
future galaxy studies to combine the most rigorous and accurate measurements of galaxy
properties with the unprecedented statistical power of upcoming observations. 

\section*{Acknowledgement}
The authors would like to thank Rita Tojeiro for her helpful comments and feedback on the draft. KJK acknowledges support from the National Science Foundation under Grant No. 2012086. 
CH is supported by the AI Accelerator program of the Schmidt Futures Foundation.
This project has received funding from the European Research Council (ERC) under the European Union’s Horizon 2020 research and innovation programme (grant agreement no. 101018897 CosmicExplorer). JA was partially supported by the research project grant “Fundamental Physics from Cosmological Surveys” funded by the Swedish Research Council (VR) under Dnr 2017- 04212.

\bibliography{sample63}{}

\begin{thebibliography}{}
\expandafter\ifx\csname natexlab\endcsname\relax\def\natexlab#1{#1}\fi
\providecommand{\url}[1]{\href{#1}{#1}}
\providecommand{\dodoi}[1]{doi:~\href{http://doi.org/#1}{\nolinkurl{#1}}}
\providecommand{\doeprint}[1]{\href{http://ascl.net/#1}{\nolinkurl{http://ascl.net/#1}}}
\providecommand{\doarXiv}[1]{\href{https://arxiv.org/abs/#1}{\nolinkurl{https://arxiv.org/abs/#1}}}

\bibitem[{Abareshi {et~al.}(2022)Abareshi, Aguilar, Ahlen, Alam, Alexander,
  Alfarsy, Allen, Prieto, Alves, Ameel, Armengaud, Asorey, Aviles, Bailey,
  Balaguera-Antolínez, Ballester, Baltay, Bault, Beltran, Benavides, BenZvi,
  Berti, Besuner, Beutler, Bianchi, Blake, Blanc, Blum, Bolton, Bose, Bramall,
  Brieden, Brodzeller, Brooks, Brownewell, Buckley-Geer, Cahn, Cai, Canning,
  Rosell, Carton, Casas, Castander, Cervantes-Cota, Chabanier, Chaussidon,
  Chuang, Circosta, Cole, Cooper, da~Costa, Cousinou, Cuceu, Davis, Dawson,
  de~la Cruz-Noriega, de~la Macorra, de~Mattia, Della~Costa, Demmer, Derwent,
  Dey, Dey, Dhungana, Ding, Dobson, Doel, Donald-McCann, Donaldson, Douglass,
  Duan, Dunlop, Edelstein, Eftekharzadeh, Eisenstein, Enriquez-Vargas,
  Escoffier, Evatt, Fagrelius, Fan, Fanning, Fawcett, Ferraro, Ereza, Flaugher,
  Font-Ribera, Forero-Romero, Frenk, Fromenteau, Gänsicke, Garcia-Quintero,
  Garrison, Gaztañaga, Gerardi, Gil-Marín, Gontcho, Gonzalez-Morales,
  Gonzalez-de Rivera, Gonzalez-Perez, Gordon, Graur, Green, Grove, Gruen,
  Gutierrez, Guy, Hahn, Harris, Herrera, Herrera-Alcantar, Honscheid, Howlett,
  Huterer, Iršič, Ishak, Jelinsky, Jiang, Jimenez, Jing, Joyce, Jullo,
  Juneau, Karaçaylı, Karamanis, Karcher, Karim, Kehoe, Kent, Kirkby, Kisner,
  Kitaura, Koposov, Kovács, Kremin, Krolewski, L'Huillier, Lahav, Lambert,
  Lamman, Lan, Landriau, Lane, Lang, Lange, Lasker, Guillou, Leauthaud,
  Van~Suu, Levi, Li, Magneville, Manera, Manser, Marshall, McCollam, McDonald,
  Meisner, Mezcua, Miller, Miquel, Montero-Camacho, Moon, Martini,
  Meneses-Rizo, Moustakas, Mueller, Muñoz-Gutiérrez, Myers, Nadathur, Najita,
  Napolitano, Neilsen, Newman, Nie, Ning, Niz, Norberg, Noriega, O'Brien,
  Obuljen, Palanque-Delabrouille, Palmese, Zhiwei, Pappalardo, Peng, Percival,
  Perruchot, Pogge, Poppett, Porredon, Prada, Prochaska, Pucha,
  Pérez-Fernández, Pérez-Ráfols, Rabinowitz, Raichoor, Ramirez-Solano,
  Ramírez-Pérez, Ravoux, Reil, Rezaie, Rocher, Rockosi, Roe, Roodman, Ross,
  Rossi, Ruggeri, Ruhlmann-Kleider, Sabiu, Safonova, Said, Saintonge, Catonga,
  Samushia, Sanchez, Saulder, Schaan, Schlafly, Schlegel, Schmoll, Scholte,
  Schubnell, Secroun, Seo, Serrano, Sharples, Sholl, Silber, Silva, Sirk,
  Siudek, Smith, Sprayberry, Staten, Stupak, Tan, Tarlé, Tie, Tojeiro,
  Ureña-López, Valdes, Valenzuela, Valluri, Vargas-Magaña, Verde, Walther,
  Wang, Wang, Weaver, Weaverdyck, Wechsler, Wilson, Yang, Yu, Yuan, Yèche,
  Zhang, Zhang, Zhao, Zhou, Zhou, Zou, Zou, Zou, \& Zu}]{desi_instrument}
Abareshi, B., Aguilar, J., Ahlen, S., {et~al.} 2022, Overview of the
  Instrumentation for the Dark Energy Spectroscopic Instrument,  arXiv,
  \dodoi{10.48550/ARXIV.2205.10939}

\bibitem[{Acquaviva {et~al.}(2011)Acquaviva, Gawiser, \&
  Guaita}]{acquaviva2011}
Acquaviva, V., Gawiser, E., \& Guaita, L. 2011, The Astrophysical Journal, 737,
  47, \dodoi{10.1088/0004-637X/737/2/47}

\bibitem[{{Alsing} {et~al.}(2022){Alsing}, {Peiris}, {Mortlock}, {Leja}, \&
  {Leistedt}}]{alsing2022}
{Alsing}, J., {Peiris}, H., {Mortlock}, D., {Leja}, J., \& {Leistedt}, B. 2022,
  arXiv e-prints, arXiv:2207.05819.
\newblock \doarXiv{2207.05819}

\bibitem[{{Alsing} {et~al.}(2020){Alsing}, {Peiris}, {Leja}, {Hahn}, {Tojeiro},
  {Mortlock}, {Leistedt}, {Johnson}, \& {Conroy}}]{speculator}
{Alsing}, J., {Peiris}, H., {Leja}, J., {et~al.} 2020, \apjs, 249, 5,
  \dodoi{10.3847/1538-4365/ab917f}

\bibitem[{Bruzual \& Charlot(2003)}]{bruzual2003}
Bruzual, G., \& Charlot, S. 2003, Monthly Notices of the Royal Astronomical
  Society, 344, 1000, \dodoi{10.1046/j.1365-8711.2003.06897.x}

\bibitem[{Carnall {et~al.}(2019)Carnall, Leja, Johnson, McLure, Dunlop, \&
  Conroy}]{carnall2019a}
Carnall, A.~C., Leja, J., Johnson, B.~D., {et~al.} 2019, The Astrophysical
  Journal, 873, 44, \dodoi{10.3847/1538-4357/ab04a2}

\bibitem[{Carnall {et~al.}(2018)Carnall, McLure, Dunlop, \&
  Davé}]{carnall2018}
Carnall, A.~C., McLure, R.~J., Dunlop, J.~S., \& Davé, R. 2018, Monthly
  Notices of the Royal Astronomical Society, 480, 4379,
  \dodoi{10.1093/mnras/sty2169}

\bibitem[{{Chabrier}(2003)}]{chabrier_2003}
{Chabrier}, G. 2003, \pasp, 115, 763, \dodoi{10.1086/376392}

\bibitem[{{Charlot} \& {Fall}(2000)}]{dust_charlot}
{Charlot}, S., \& {Fall}, S.~M. 2000, \apj, 539, 718, \dodoi{10.1086/309250}

\bibitem[{Chevallard \& Charlot(2016)}]{chevallard2016}
Chevallard, J., \& Charlot, S. 2016, Monthly Notices of the Royal Astronomical
  Society, 462, 1415, \dodoi{10.1093/mnras/stw1756}

\bibitem[{{Choi} {et~al.}(2016){Choi}, {Dotter}, {Conroy}, {Cantiello},
  {Paxton}, \& {Johnson}}]{mist_2}
{Choi}, J., {Dotter}, A., {Conroy}, C., {et~al.} 2016, \apj, 823, 102,
  \dodoi{10.3847/0004-637X/823/2/102}

\bibitem[{Cichocki \& Phan(2009)}]{nmf2}
Cichocki, A., \& Phan, A.-H. 2009, IEICE Transactions on Fundamentals of
  Electronics, Communications and Computer Sciences, E92.A, 708,
  \dodoi{10.1587/transfun.E92.A.708}

\bibitem[{Conroy(2013)}]{conroy2013}
Conroy, C. 2013, Annual Review of Astronomy and Astrophysics, 51, 393,
  \dodoi{10.1146/annurev-astro-082812-141017}

\bibitem[{{Conroy} \& {Gunn}(2010)}]{fsps_model}
{Conroy}, C., \& {Gunn}, J.~E. 2010, \apj, 712, 833,
  \dodoi{10.1088/0004-637X/712/2/833}

\bibitem[{Conroy {et~al.}(2009)Conroy, Gunn, \& White}]{conroy2009}
Conroy, C., Gunn, J.~E., \& White, M. 2009, The Astrophysical Journal, 699,
  486, \dodoi{10.1088/0004-637X/699/1/486}

\bibitem[{{Dalton} {et~al.}(2012){Dalton}, {Trager}, {Abrams}, {Carter},
  {Bonifacio}, {Aguerri}, {MacIntosh}, {Evans}, {Lewis}, {Navarro}, {Agocs},
  {Dee}, {Rousset}, {Tosh}, {Middleton}, {Pragt}, {Terrett}, {Brock}, {Benn},
  {Verheijen}, {Cano Infantes}, {Bevil}, {Steele}, {Mottram}, {Bates},
  {Gribbin}, {Rey}, {Rodriguez}, {Delgado}, {Guinouard}, {Walton}, {Irwin},
  {Jagourel}, {Stuik}, {Gerlofsma}, {Roelfsma}, {Skillen}, {Ridings},
  {Balcells}, {Daban}, {Gouvret}, {Venema}, \& {Girard}}]{weave}
{Dalton}, G., {Trager}, S.~C., {Abrams}, D.~C., {et~al.} 2012, in Society of
  Photo-Optical Instrumentation Engineers (SPIE) Conference Series, Vol. 8446,
  Ground-based and Airborne Instrumentation for Astronomy IV, ed. I.~S.
  {McLean}, S.~K. {Ramsay}, \& H.~{Takami}, 84460P, \dodoi{10.1117/12.925950}

\bibitem[{{de Jong} {et~al.}(2019){de Jong}, {Agertz}, {Berbel}, {Aird},
  {Alexander}, {Amarsi}, {Anders}, {Andrae}, {Ansarinejad}, {Ansorge},
  {Antilogus}, {Anwand-Heerwart}, {Arentsen}, {Arnadottir}, {Asplund}, {Auger},
  {Azais}, {Baade}, {Baker}, {Baker}, {Balbinot}, {Baldry}, {Banerji},
  {Barden}, {Barklem}, {Barth{\'e}l{\'e}my-Mazot}, {Battistini}, {Bauer},
  {Bell}, {Bellido-Tirado}, {Bellstedt}, {Belokurov}, {Bensby}, {Bergemann},
  {Bestenlehner}, {Bielby}, {Bilicki}, {Blake}, {Bland-Hawthorn}, {Boeche},
  {Boland}, {Boller}, {Bongard}, {Bongiorno}, {Bonifacio}, {Boudon}, {Brooks},
  {Brown}, {Brown}, {Br{\"u}ggen}, {Brynnel}, {Brzeski}, {Buchert},
  {Buschkamp}, {Caffau}, {Caillier}, {Carrick}, {Casagrande}, {Case}, {Casey},
  {Cesarini}, {Cescutti}, {Chapuis}, {Chiappini}, {Childress}, {Christlieb},
  {Church}, {Cioni}, {Cluver}, {Colless}, {Collett}, {Comparat}, {Cooper},
  {Couch}, {Courbin}, {Croom}, {Croton}, {Daguis{\'e}}, {Dalton}, {Davies},
  {Davis}, {de Laverny}, {Deason}, {Dionies}, {Disseau}, {Doel}, {D{\"o}scher},
  {Driver}, {Dwelly}, {Eckert}, {Edge}, {Edvardsson}, {Youssoufi}, {Elhaddad},
  {Enke}, {Erfanianfar}, {Farrell}, {Fechner}, {Feiz}, {Feltzing}, {Ferreras},
  {Feuerstein}, {Feuillet}, {Finoguenov}, {Ford}, {Fotopoulou}, {Fouesneau},
  {Frenk}, {Frey}, {Gaessler}, {Geier}, {Gentile Fusillo}, {Gerhard},
  {Giannantonio}, {Giannone}, {Gibson}, {Gillingham},
  {Gonz{\'a}lez-Fern{\'a}ndez}, {Gonzalez-Solares}, {Gottloeber}, {Gould},
  {Grebel}, {Gueguen}, {Guiglion}, {Haehnelt}, {Hahn}, {Hansen}, {Hartman},
  {Hauptner}, {Hawkins}, {Haynes}, {Haynes}, {Heiter}, {Helmi}, {Aguayo},
  {Hewett}, {Hinton}, {Hobbs}, {Hoenig}, {Hofman}, {Hook}, {Hopgood},
  {Hopkins}, {Hourihane}, {Howes}, {Howlett}, {Huet}, {Irwin}, {Iwert},
  {Jablonka}, {Jahn}, {Jahnke}, {Jarno}, {Jin}, {Jofre}, {Johl}, {Jones},
  {J{\"o}nsson}, {Jordan}, {Karovicova}, {Khalatyan}, {Kelz}, {Kennicutt},
  {King}, {Kitaura}, {Klar}, {Klauser}, {Kneib}, {Koch}, {Koposov},
  {Kordopatis}, {Korn}, {Kosmalski}, {Kotak}, {Kovalev}, {Kreckel}, {Kripak},
  {Krumpe}, {Kuijken}, {Kunder}, {Kushniruk}, {Lam}, {Lamer}, {Laurent},
  {Lawrence}, {Lehmitz}, {Lemasle}, {Lewis}, {Li}, {Lidman}, {Lind}, {Liske},
  {Lizon}, {Loveday}, {Ludwig}, {McDermid}, {Maguire}, {Mainieri}, {Mali},
  {Mandel}, {Mandel}, {Mannering}, {Martell}, {Martinez Delgado}, {Matijevic},
  {McGregor}, {McMahon}, {McMillan}, {Mena}, {Merloni}, {Meyer}, {Michel},
  {Micheva}, {Migniau}, {Minchev}, {Monari}, {Muller}, {Murphy},
  {Muthukrishna}, {Nandra}, {Navarro}, {Ness}, {Nichani}, {Nichol}, {Nicklas},
  {Niederhofer}, {Norberg}, {Obreschkow}, {Oliver}, {Owers}, {Pai},
  {Pankratow}, {Parkinson}, {Paschke}, {Paterson}, {Pecontal}, {Parry},
  {Phillips}, {Pillepich}, {Pinard}, {Pirard}, {Piskunov}, {Plank},
  {Pl{\"u}schke}, {Pons}, {Popesso}, {Power}, {Pragt}, {Pramskiy}, {Pryer},
  {Quattri}, {Queiroz}, {Quirrenbach}, {Rahurkar}, {Raichoor}, {Ramstedt},
  {Rau}, {Recio-Blanco}, {Reiss}, {Renaud}, {Revaz}, {Rhode}, {Richard},
  {Richter}, {Rix}, {Robotham}, {Roelfsema}, {Romaniello}, {Rosario},
  {Rothmaier}, {Roukema}, {Ruchti}, {Rupprecht}, {Rybizki}, {Ryde}, {Saar},
  {Sadler}, {Sahl{\'e}n}, {Salvato}, {Sassolas}, {Saunders}, {Saviauk},
  {Sbordone}, {Schmidt}, {Schnurr}, {Scholz}, {Schwope}, {Seifert}, {Shanks},
  {Sheinis}, {Sivov}, {Sk{\'u}lad{\'o}ttir}, {Smartt}, {Smedley}, {Smith},
  {Smith}, {Sorce}, {Spitler}, {Starkenburg}, {Steinmetz}, {Stilz}, {Storm},
  {Sullivan}, {Sutherland}, {Swann}, {Tamone}, {Taylor}, {Teillon}, {Tempel},
  {ter Horst}, {Thi}, {Tolstoy}, {Trager}, {Traven}, {Tremblay}, {Tresse},
  {Valentini}, {van de Weygaert}, {van den Ancker}, {Veljanoski}, {Venkatesan},
  {Wagner}, {Wagner}, {Walcher}, {Waller}, {Walton}, {Wang}, {Winkler},
  {Wisotzki}, {Worley}, {Worseck}, {Xiang}, {Xu}, {Yong}, {Zhao}, {Zheng},
  {Zscheyge}, \& {Zucker}}]{4most}
{de Jong}, R.~S., {Agertz}, O., {Berbel}, A.~A., {et~al.} 2019, The Messenger,
  175, 3, \dodoi{10.18727/0722-6691/5117}

\bibitem[{{DESI Collaboration} {et~al.}(2016){DESI Collaboration}, Aghamousa,
  Aguilar, Ahlen, Alam, Allen, Prieto, Annis, Bailey, Balland, Ballester,
  Baltay, Beaufore, Bebek, Beers, Bell, Bernal, Besuner, Beutler, Blake,
  Bleuler, Blomqvist, Blum, Bolton, Briceno, Brooks, Brownstein, Buckley-Geer,
  Burden, Burtin, Busca, Cahn, Cai, Cardiel-Sas, Carlberg, Carton, Casas,
  Castander, Cervantes-Cota, Claybaugh, Close, Coker, Cole, Comparat, Cooper,
  Cousinou, Crocce, Cuby, Cunningham, Davis, Dawson, de~la Macorra, De~Vicente,
  Delubac, Derwent, Dey, Dhungana, Ding, Doel, Duan, Ealet, Edelstein,
  Eftekharzadeh, Eisenstein, Elliott, Escoffier, Evatt, Fagrelius, Fan,
  Fanning, Farahi, Farihi, Favole, Feng, Fernandez, Findlay, Finkbeiner,
  Fitzpatrick, Flaugher, Flender, Font-Ribera, Forero-Romero, Fosalba, Frenk,
  Fumagalli, Gaensicke, Gallo, Garcia-Bellido, Gaztanaga, Fusillo, Gerard,
  Gershkovich, Giannantonio, Gillet, Gonzalez-de Rivera, Gonzalez-Perez, Gott,
  Graur, Gutierrez, Guy, Habib, Heetderks, Heetderks, Heitmann, Hellwing,
  Herrera, Ho, Holland, Honscheid, Huff, Hutchinson, Huterer, Hwang, Laguna,
  Ishikawa, Jacobs, Jeffrey, Jelinsky, Jennings, Jiang, Jimenez, Johnson,
  Joyce, Jullo, Juneau, Kama, Karcher, Karkar, Kehoe, Kennamer, Kent,
  Kilbinger, Kim, Kirkby, Kisner, Kitanidis, Kneib, Koposov, Kovacs, Koyama,
  Kremin, Kron, Kronig, Kueter-Young, Lacey, Lafever, Lahav, Lambert, Lampton,
  Landriau, Lang, Lauer, Goff, Guillou, Van~Suu, Lee, Lee, Leitner, Lesser,
  Levi, L'Huillier, Li, Liang, Lin, Linder, Loebman, Lukić, Ma, MacCrann,
  Magneville, Makarem, Manera, Manser, Marshall, Martini, Massey, Matheson,
  McCauley, McDonald, McGreer, Meisner, Metcalfe, Miller, Miquel, Moustakas,
  Myers, Naik, Newman, Nichol, Nicola, da~Costa, Nie, Niz, Norberg, Nord,
  Norman, Nugent, O'Brien, Oh, Olsen, Padilla, Padmanabhan, Padmanabhan,
  Palanque-Delabrouille, Palmese, Pappalardo, Pâris, Park, Patej, Peacock,
  Peiris, Peng, Percival, Perruchot, Pieri, Pogge, Pollack, Poppett, Prada,
  Prakash, Probst, Rabinowitz, Raichoor, Ree, Refregier, Regal, Reid, Reil,
  Rezaie, Rockosi, Roe, Ronayette, Roodman, Ross, Ross, Rossi, Rozo,
  Ruhlmann-Kleider, Rykoff, Sabiu, Samushia, Sanchez, Sanchez, Schlegel,
  Schneider, Schubnell, Secroun, Seljak, Seo, Serrano, Shafieloo, Shan,
  Sharples, Sholl, Shourt, Silber, Silva, Sirk, Slosar, Smith, Smoot, Som,
  Song, Sprayberry, Staten, Stefanik, Tarle, Tie, Tinker, Tojeiro, Valdes,
  Valenzuela, Valluri, Vargas-Magana, Verde, Walker, Wang, Wang, Weaver,
  Weaverdyck, Wechsler, Weinberg, White, Yang, Yeche, Zhang, Zhao, Zheng, Zhou,
  Zhou, Zhu, Zou, \& Zu}]{desicollaboration2016}
{DESI Collaboration}, Aghamousa, A., Aguilar, J., {et~al.} 2016,
  arXiv:1611.00036 [astro-ph].
\newblock \url{http://arxiv.org/abs/1611.00036}

\bibitem[{{Dotter}(2016)}]{mist_1}
{Dotter}, A. 2016, \apjs, 222, 8, \dodoi{10.3847/0067-0049/222/1/8}

\bibitem[{Foreman-Mackey {et~al.}(2014)Foreman-Mackey, Sick, \&
  Johnson}]{python_fsps}
Foreman-Mackey, D., Sick, J., \& Johnson, B. 2014, python-fsps: Python bindings
  to FSPS (v0.1.1), v0.1.1,  Zenodo, \dodoi{10.5281/zenodo.12157}

\bibitem[{Févotte \& Idier(2011)}]{nmf3}
Févotte, C., \& Idier, J. 2011, Algorithms for nonnegative matrix
  factorization with the beta-divergence.
\newblock \doarXiv{1010.1763}

\bibitem[{{Goodman} \& {Weare}(2010)}]{gw}
{Goodman}, J., \& {Weare}, J. 2010, Communications in Applied Mathematics and
  Computational Science, 5, 65, \dodoi{10.2140/camcos.2010.5.65}

\bibitem[{{Hahn} {et~al.}(2022{\natexlab{a}}){Hahn}, {Kwon}, {Tojeiro},
  {Siudek}, {Canning}, {Mezcua}, {Tinker}, {Brooks}, {Doel}, {Fanning},
  {Gazta{\~n}aga}, {Kehoe}, {Landriau}, {Meisner}, {Moustakas}, {Poppett},
  {Tarle}, {Weiner}, \& {Zou}}]{provabgs}
{Hahn}, C., {Kwon}, K.~J., {Tojeiro}, R., {et~al.} 2022{\natexlab{a}}, arXiv
  e-prints, arXiv:2202.01809.
\newblock \doarXiv{2202.01809}

\bibitem[{{Hahn} {et~al.}(2022{\natexlab{b}}){Hahn}, {Wilson}, {Ruiz-Macias},
  {Cole}, {Weinberg}, {Moustakas}, {Kremin}, {Tinker}, {Smith}, {Wechsler},
  {Ahlen}, {Alam}, {Bailey}, {Brooks}, {Cooper}, {Davis}, {Dawson}, {Dey},
  {Dey}, {Eftekharzadeh}, {Eisenstein}, {Fanning}, {Forero-Romero}, {Frenk},
  {Gazta{\~n}aga}, {Gontcho}, {Guy}, {Honscheid}, {Ishak}, {Juneau}, {Kehoe},
  {Kisner}, {Lan}, {Landriau}, {Le Guillou}, {Levi}, {Magneville}, {Martini},
  {Meisner}, {Myers}, {Nie}, {Norberg}, {Palanque-Delabrouille}, {Percival},
  {Poppett}, {Prada}, {Raichoor}, {Ross}, {Safonova}, {Saulder}, {Schlafly},
  {Schlegel}, {Sierra-Porta}, {Tarle}, {Weaver}, {Y{\`e}che}, {Zarrouk},
  {Zhou}, {Zhou}, \& {Zou}}]{hahn2022}
{Hahn}, C., {Wilson}, M.~J., {Ruiz-Macias}, O., {et~al.} 2022{\natexlab{b}},
  arXiv e-prints, arXiv:2208.08512.
\newblock \doarXiv{2208.08512}

\bibitem[{Henriques {et~al.}(2015)Henriques, White, Thomas, Angulo, Guo,
  Lemson, Springel, \& Overzier}]{lgal}
Henriques, B. M.~B., White, S. D.~M., Thomas, P.~A., {et~al.} 2015, Monthly
  Notices of the Royal Astronomical Society, 451, 2663,
  \dodoi{10.1093/mnras/stv705}

\bibitem[{Johnson {et~al.}(2021)Johnson, Leja, Conroy, \&
  Speagle}]{johnson2021}
Johnson, B.~D., Leja, J., Conroy, C., \& Speagle, J.~S. 2021, The Astrophysical
  Journal Supplement Series, 254, 22, \dodoi{10.3847/1538-4365/abef67}

\bibitem[{Karamanis \& Beutler(2020)}]{zeus2}
Karamanis, M., \& Beutler, F. 2020, arXiv preprint arXiv: 2002.06212

\bibitem[{Karamanis {et~al.}(2021)Karamanis, Beutler, \& Peacock}]{zeus1}
Karamanis, M., Beutler, F., \& Peacock, J.~A. 2021, arXiv preprint
  arXiv:2105.03468

\bibitem[{{Kingma} \& {Ba}(2014)}]{adam}
{Kingma}, D.~P., \& {Ba}, J. 2014, arXiv e-prints, arXiv:1412.6980.
\newblock \doarXiv{1412.6980}

\bibitem[{{Kriek} \& {Conroy}(2013)}]{kriek_conroy}
{Kriek}, M., \& {Conroy}, C. 2013, \apjl, 775, L16,
  \dodoi{10.1088/2041-8205/775/1/L16}

\bibitem[{Lee \& Seung(1999)}]{nmf1}
Lee, D.~D., \& Seung, H.~S. 1999, Nature, 401, 788, \dodoi{10.1038/44565}

\bibitem[{{Leistedt} {et~al.}(2022){Leistedt}, {Alsing}, {Peiris}, {Mortlock},
  \& {Leja}}]{leistedt2022}
{Leistedt}, B., {Alsing}, J., {Peiris}, H., {Mortlock}, D., \& {Leja}, J. 2022,
  arXiv e-prints, arXiv:2207.07673.
\newblock \doarXiv{2207.07673}

\bibitem[{Leja {et~al.}(2017)Leja, Johnson, Conroy, van Dokkum, \&
  Byler}]{leja2017}
Leja, J., Johnson, B.~D., Conroy, C., van Dokkum, P.~G., \& Byler, N. 2017, The
  Astrophysical Journal, 837, 170, \dodoi{10.3847/1538-4357/aa5ffe}

\bibitem[{Leja {et~al.}(2019)Leja, Speagle, Johnson, Conroy, van Dokkum, \&
  Franx}]{leja2019a}
Leja, J., Speagle, J.~S., Johnson, B.~D., {et~al.} 2019, arXiv,
  arXiv:1910.04168.
\newblock \url{https://ui.adsabs.harvard.edu/abs/2019arXiv191004168L/abstract}

\bibitem[{{Leja} {et~al.}(2019){Leja}, {Johnson}, {Conroy}, {van Dokkum},
  {Speagle}, {Brammer}, {Momcheva}, {Skelton}, {Whitaker}, {Franx}, \&
  {Nelson}}]{Leja+_19}
{Leja}, J., {Johnson}, B.~D., {Conroy}, C., {et~al.} 2019, \apj, 877, 140,
  \dodoi{10.3847/1538-4357/ab1d5a}

\bibitem[{{Leja} {et~al.}(2021){Leja}, {Speagle}, {Ting}, {Johnson}, {Conroy},
  {Whitaker}, {Nelson}, {van Dokkum}, \& {Franx}}]{leja2021b}
{Leja}, J., {Speagle}, J.~S., {Ting}, Y.-S., {et~al.} 2021, arXiv e-prints,
  arXiv:2110.04314.
\newblock \doarXiv{2110.04314}

\bibitem[{{Lejeune} {et~al.}(1997){Lejeune}, {Cuisinier}, \& {Buser}}]{basel1}
{Lejeune}, T., {Cuisinier}, F., \& {Buser}, R. 1997, \aaps, 125, 229,
  \dodoi{10.1051/aas:1997373}

\bibitem[{{Lejeune} {et~al.}(1998){Lejeune}, {Cuisinier}, \& {Buser}}]{basel2}
---. 1998, \aaps, 130, 65, \dodoi{10.1051/aas:1998405}

\bibitem[{Maraston(2005)}]{maraston2005}
Maraston, C. 2005, Monthly Notices of the Royal Astronomical Society, 362, 799,
  \dodoi{10.1111/j.1365-2966.2005.09270.x}

\bibitem[{{Nelson} {et~al.}(2015){Nelson}, {Pillepich}, {Genel},
  {Vogelsberger}, {Springel}, {Torrey}, {Rodriguez-Gomez}, {Sijacki}, {Snyder},
  {Griffen}, {Marinacci}, {Blecha}, {Sales}, {Xu}, \& {Hernquist}}]{illustris1}
{Nelson}, D., {Pillepich}, A., {Genel}, S., {et~al.} 2015, Astronomy and
  Computing, 13, 12, \dodoi{10.1016/j.ascom.2015.09.003}

\bibitem[{{Paxton} {et~al.}(2011){Paxton}, {Bildsten}, {Dotter}, {Herwig},
  {Lesaffre}, \& {Timmes}}]{mist_3.1}
{Paxton}, B., {Bildsten}, L., {Dotter}, A., {et~al.} 2011, \apjs, 192, 3,
  \dodoi{10.1088/0067-0049/192/1/3}

\bibitem[{{Paxton} {et~al.}(2013){Paxton}, {Cantiello}, {Arras}, {Bildsten},
  {Brown}, {Dotter}, {Mankovich}, {Montgomery}, {Stello}, {Timmes}, \&
  {Townsend}}]{mist_3.2}
{Paxton}, B., {Cantiello}, M., {Arras}, P., {et~al.} 2013, \apjs, 208, 4,
  \dodoi{10.1088/0067-0049/208/1/4}

\bibitem[{{Paxton} {et~al.}(2015){Paxton}, {Marchant}, {Schwab}, {Bauer},
  {Bildsten}, {Cantiello}, {Dessart}, {Farmer}, {Hu}, {Langer}, {Townsend},
  {Townsley}, \& {Timmes}}]{mist_3.3}
{Paxton}, B., {Marchant}, P., {Schwab}, J., {et~al.} 2015, \apjs, 220, 15,
  \dodoi{10.1088/0067-0049/220/1/15}

\bibitem[{Sánchez-Blázquez {et~al.}(2006)Sánchez-Blázquez, Peletier,
  Jiménez-Vicente, Cardiel, Cenarro, Falcón-Barroso, Gorgas, Selam, \&
  Vazdekis}]{miles}
Sánchez-Blázquez, P., Peletier, R.~F., Jiménez-Vicente, J., {et~al.} 2006,
  Monthly Notices of the Royal Astronomical Society, 371, 703,
  \dodoi{10.1111/j.1365-2966.2006.10699.x}

\bibitem[{Tacchella {et~al.}(2021)Tacchella, Conroy, Faber, Johnson, Leja,
  Barro, Cunningham, Deason, Guhathakurta, Guo, Hernquist, Koo, McKinnon,
  Rockosi, Speagle, van Dokkum, \& Yesuf}]{tacchella2021}
Tacchella, S., Conroy, C., Faber, S.~M., {et~al.} 2021, arXiv e-prints, 2102,
  arXiv:2102.12494.
\newblock \url{http://adsabs.harvard.edu/abs/2021arXiv210212494T}

\bibitem[{Takada {et~al.}(2014)Takada, Ellis, Chiba, Greene, Aihara, Arimoto,
  Bundy, Cohen, Doré, Graves, Gunn, Heckman, Hirata, Ho, Kneib, Le~Fèvre,
  Lin, More, Murayama, Nagao, Ouchi, Seiffert, Silverman, Sodré, Spergel,
  Strauss, Sugai, Suto, Takami, \& Wyse}]{takada2014}
Takada, M., Ellis, R.~S., Chiba, M., {et~al.} 2014, Publications of the
  Astronomical Society of Japan, 66, R1, \dodoi{10.1093/pasj/pst019}

\bibitem[{{Vogelsberger} {et~al.}(2014){Vogelsberger}, {Genel}, {Springel},
  {Torrey}, {Sijacki}, {Xu}, {Snyder}, {Bird}, {Nelson}, \&
  {Hernquist}}]{illustris2}
{Vogelsberger}, M., {Genel}, S., {Springel}, V., {et~al.} 2014, \nat, 509, 177,
  \dodoi{10.1038/nature13316}

\bibitem[{Walcher {et~al.}(2011)Walcher, Groves, Budavári, \&
  Dale}]{walcher2011}
Walcher, J., Groves, B., Budavári, T., \& Dale, D. 2011, Astrophysics and
  Space Science, 331, 1, \dodoi{10.1007/s10509-010-0458-z}

\bibitem[{{Westera} {et~al.}(2002){Westera}, {Lejeune}, {Buser}, {Cuisinier},
  \& {Bruzual}}]{basel3}
{Westera}, P., {Lejeune}, T., {Buser}, R., {Cuisinier}, F., \& {Bruzual}, G.
  2002, \aap, 381, 524, \dodoi{10.1051/0004-6361:20011493}

\end{thebibliography}
\bibliographystyle{aasjournal}
\end{document}